\documentclass{jjap3}
\usepackage{txfonts}
\usepackage{multirow}

\title{First-principles study on various point defects formed by hydrogen and helium atoms in tungsten}
\author{Qiang Zhao$^{1}$\thanks{Corresponding author. E-mail: qzhao@ncepu.edu.cn}, Zheng Zhang$^{1}$, Yang Li$^{1}$, and Xiaoping Ouyang$^{1,2,3}$}
\inst{$^{1}$Beijing Key Laboratory of Passive Safety Technology for Nuclear Energy, School of Nuclear Science and Engineering, North China Electric Power University, Beijing 102206, P.R. China\\ $^{2}$Northwest Institute of Nuclear Technology, Xi'an 710024, P.R. China\\
$^{3}$School of Materials Science and Engineering, Xiangtan University, Xiangtan 411105, P.R. China}
\abst{The different point defects formed by two hydrogen atoms or two helium atoms in tungsten were investigated through first-principles calculation. The energetically favorable site for a hydrogen atom is tetrahedral interstitial site while substitutional site is the most preferred site for a helium atom. The formation energies of two hydrogen or helium atoms are determined by their positions, and there are not a multiple of a single atom's. After relaxation, two adjacent hydrogen atoms are away from each other while helium atoms are close to. The reasons for the interaction between two hydrogen or helium atoms are also discussed.
}
\begin{document}
\maketitle
\section{Introduction}
Tungsten (W) is widely considered as the most potential plasma facing material (PFM) \cite{lee2009energetics,skinner2008recent} and the first wall material (FWM) in fusion reactor due to its excellent properties, such as high melting point, high mechanical strength at high temperatures, efficient thermal conductivity, low thermal expansion coefficient, and high sputtering threshold energy\cite{lassner2012tungsten}. Studies on tungsten receive an increasing amount of attention from scientists and engineers \cite{guerrero2016first,kong2014first,kong2015first,xu2009first,nguyen2015trapping,alkhamees2009first,kong2012first}. Tungsten is subjected to high particle fluxes in fusion reactors, not only the plasma background ions (H, D and T) but also the intrinsic impurities (He, Be and C)\cite{federici1999vessel,federici2001assessment}. Because hydrogen is one of the major plasma background ions, hydrogen (H) irradiation may result in the change of the mechanical properties of tungsten. Helium (He) irradiation leads to the blister formation and the subsequent degradation of the mechanical properties of metals \cite{xu2007accumulation,das1976radiation,tokunaga2007effects,lee2007hydrogen,lee2007hydroge,katayama2007helium,debelle2007helium}. Thus, it is very important to investigate the trapping and blister formation of hydrogen or helium in tungsten and the behaviors of hydrogen and helium in tungsten.

A number of experimental and theoretical studies were carried out to understand
the interaction between the hydrogen or helium impurity and the tungsten host lattice \cite{lee2009energetics,nguyen2015trapping,xu2009first,arkhipov2007deuterium,ye2003blister,grigorev2015nucleation,liu2009structure,wang2015first,zhou2016modeling,wang2015comparison,guterl2015modeling,li2014molecular,grigorev2016mobility}. For example, Arkhipov et al. found that hydrogen could form a blister on the tungsten surface \cite{arkhipov2007deuterium}. Ye et al. presented the experimental results in a divertor plasma simulator NAGDIS-\mbox{II} and observed the formation of hydrogen bubbles on the surface of tungsten\cite{ye2003blister}. Grigorev et al. observed the nucleation and growth of hydrogen bubbles on dislocations in tungsten under high flux, low energy plasma exposure \cite{grigorev2015nucleation}. Lee, Xu, and Liu et al. studied the behavior of hydrogen in tungsten through first-principles calculation \cite{lee2009energetics,xu2009first,liu2009structure}. Wang, Zhou, and Wang et al. studied the behavior of helium on tungsten surfaces and the helium in tungsten by using first-principles calculation\cite{wang2015first,zhou2016modeling,wang2015comparison}. Guter, Nguyen, Li, and Ggrigorev et al. studied the behavior of hydrogen or helium in tungsten by using the molecular dynamics simulation method \cite{nguyen2015trapping,guterl2015modeling,li2014molecular,grigorev2016mobility}. With the rapid development of the computer technique, computer simulation, such us first-principles calculation\cite{Kasamatsu2009Comparative,Tani2011First,Konabe2013High,Absor2014Tunable,Iizuka2015First}, is now playing important role in the materials research.

\begin{figure}
\includegraphics[width=0.75\textwidth]{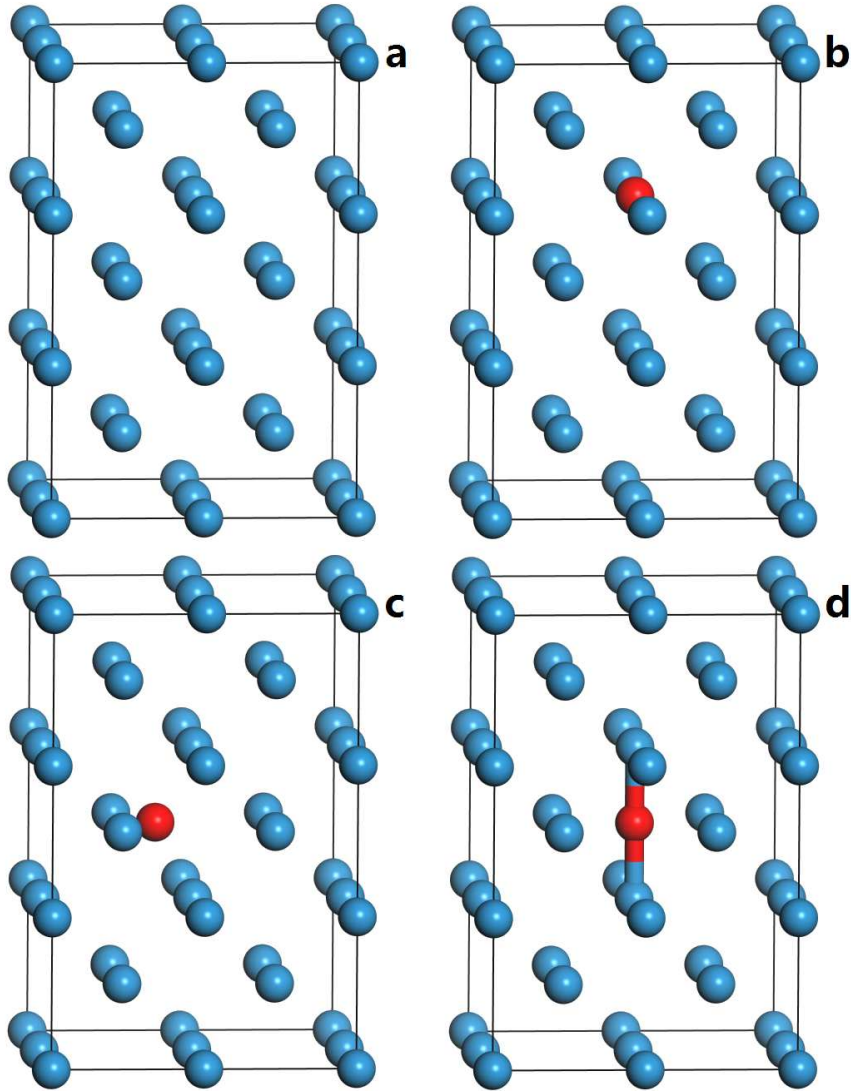}
\caption{The crystal structure of BCC W. The blue balls stand for the tungsten atoms, and the red ball stand for the H atom or He atom. (a) shows the pure W supercell, and (b), (c), and (d) show the W supercell contain a H atom or He atom in the substitutional site, tetrahedral site and octahedral site, respectively.}
\label{fig:1}
\end{figure}

There are a lot of work focus on defects in tungsten. For example, research works on the interaction between two hydrogen that are located in tetrahedral sites in tungsten have been completed\cite{xu2009first,liu2009structure}. Some researchers investigated the interaction between hydrogen or helium atoms and the vacancy\cite{lee2009energetics}. However, a systematic study of the interaction between two hydrogen or helium atoms in tungsten without other defects is still lacked. In this paper, by using first-principles calculation based on density functional theory (DFT), we have studied the behavior of hydrogen atoms and helium atoms in body centered cubic (BCC) tungsten. We calculated the formation energy of single hydrogen atom and two hydrogen atoms in tungsten and the formation energy of single helium atom and two helium atoms in tungsten. Using the charge density difference and atomic population, we analyzed the reason why there is a repulsive interaction between two hydrogen atoms and an attractive interaction between two helium atoms. These computational results might shed some light on the development of tungsten-based materials that are used in ITER and other fusion rectors.
\section{Models and computational details}
\begin{figure}
\includegraphics[width=0.75\textwidth]{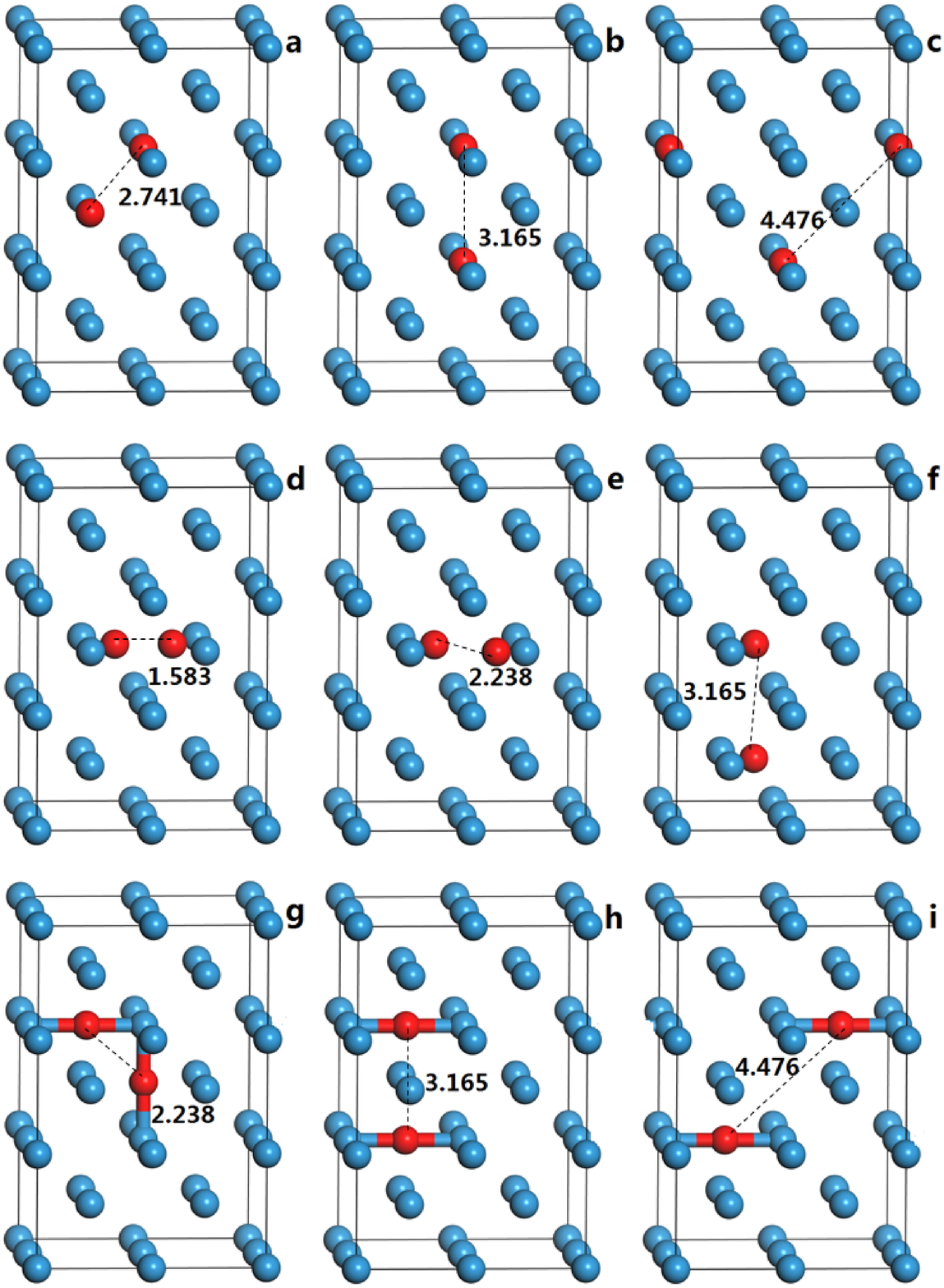}
\caption{The crystal structure of BCC W contains two H atoms or He atoms. The blue balls and red balls stand for W atoms, and the H atoms or He atoms, respectively. (a), (b), and (c) show two atoms located in the substitutional site. (d), (e), and (f) show two atoms located in the tetrahedral site, and (g), (h), and (i) show two atoms located in the octahedral site. The numbers in the figure stand for the distances (in \AA) of between the two foreign atoms.}
\label{fig.2}
\end{figure}
Tungsten belongs to the BCC crystal system. Its space group is Im-3m. The simulation supercell composed of 24 lattice points ($2\times3\times2$) was used in this paper. Initially, we considered three possible sites for a single H atom and He atom enter the lattice of BCC W. According to their locations, we marked the H and He atoms as H$_{\rm{Sub}}$, H$_{\rm{Tet}}$, H$_{\rm{Oct}}$, He$_{\rm{Sub}}$, He$_{\rm{Tet}}$, and He$_{\rm{Oct}}$ when the H atom and He atom were located in the substitutional site, the tetrahedral interstitial site, and the octahedral interstitial site, respectively. The calculation models that contain a single H atom or a He atom are shown in Fig. 1. To explore the interaction between two H atoms or two He atoms, we built a series of two H atoms or two He atoms models. The symbols H$_{\rm{Sub}}^{\rm{21}}$, H$_{\rm{Sub}}^{\rm{22}}$, and H$_{\rm{Sub}}^{\rm{23}}$ stand for the first-nearest-neighbor, second-nearest-neighbor, and third-nearest-neighbor H atom around the pre-existing H atom that is located in the substitutional site, respectively. The symbols H$_{\rm{Tet}}^{\rm{21}}$, H$_{\rm{Tet}}^{\rm{22}}$, H$_{\rm{Tet}}^{\rm{23}}$, H$_{\rm{Oct}}^{\rm{21}}$, H$_{\rm{Oct}}^{\rm{22}}$, and H$_{\rm{Oct}}^{\rm{23}}$ stand for the first-nearest-neighbor, second-nearest-neighbor, and third-nearest-neighbor H atom around the pre-existing H atom that is located in tetrahedral and octahedral site, respectively. We use the symbols He$_{\rm{Sub}}^{\rm{21}}$, He$_{\rm{Sub}}^{\rm{22}}$, He$_{\rm{Sub}}^{\rm{23}}$, He$_{\rm{Tet}}^{\rm{21}}$, He$_{\rm{Tet}}^{\rm{22}}$, He$_{\rm{Tet}}^{\rm{23}}$, He$_{\rm{Oct}}^{\rm{21}}$, He$_{\rm{Oct}}^{\rm{22}}$, and He$_{\rm{Oct}}^{\rm{23}}$ stand for the first-nearest-neighbor, second-nearest-neighbor, and third-nearest-neighbor He around the pre-existing He atom that is located in substitutional, tetrahedral, and octahedral sites, respectively. These models contain two H atoms or two He atoms and are shown in Fig. 2.

First-principles calculations were performed by using the density functional theory (DFT) \cite{hohenberg1964inhomogeneous,kohn1965self} and the plane-wave pseudo-potential technique, as implemented in the Cambridge Squential Total Energy Package (CASTEP) \cite{clark2005first}. The generalized gradient approximation (GGA) \cite{perdew1986accurate,perdew1992atoms,perdew1996generalized,iikura2001long} with Perdew-Burke-Ernzerhof (PBE) \cite{perdew1996generalized} functional was used to describe the exchange-correlation interaction among the electrons, and ultrasoft pseudo-potentials were employed for ion-electron interaction. For W, the four 5$d$ electrons are considered to be the valence electrons together with two 6$s$ electrons (the reference state is $5d^46s^2$). During the calculation, the supercell size, shape, and atomic position are relaxed to equilibrium. The plane-wave energy cutoff is 400 eV for all calculations. The basic parameters were chosen as follows: the space representation is reciprocal, SCF tolerance equals $1.0\times10^{-6}$ eV/atom, and $k$ sampling with $5\times5\times5$ $k$-point mesh in the Brillouin zone was used. After the convergence test was completed, the calculation parameters were chosen as follows: (1) the maximum inter-atomic forces were smaller than 0.05 eV/nm; (2) the maximum change of energy per atom was smaller than $1.0\times10^{-5}$ eV; (3) the maximum displacement was smaller than 0.001\AA; and (4) the maximum stress of the crystal was smaller than 0.02 GPa. All the properties that we calculated based on the crystal structure have been optimized.
\begin{figure}
  \centering
  \includegraphics[width=0.75\textwidth]{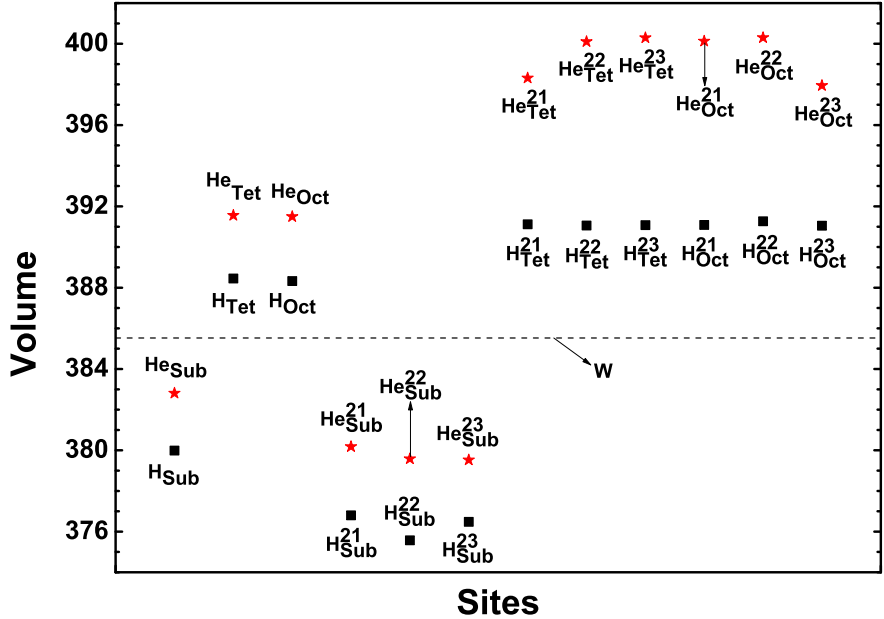}
  \caption{The volume (in \AA$^3$) of the supercell ($2\times3\times2$) of the W supercell with single and two point defects formed by H and He atoms. The dashed line represents the volume of the perfect BCC W supercell.}
  \label{fig.3}
\end{figure}

\begin{table}
\caption{Calculated and experimental lattice constants (a in \AA), elastic constants
($C_{ij}$ in GPa), bulk modulus ($B$ in GPa), Young's modulus ($E$ in GPa), shear modulus ($G$ in GPa), Poisson's ratio ($\nu$), and $G/B$ for BCC W.}
\label{t1}
\begin{tabular}{cccccccccc}
\Hline
Parameters& a& $C_{11}$& $C_{12}$& $C_{44}$ & $B$ & $E$ & $G$ & $\nu$ & $G/B$\\
\Hline
This paper & 3.18 & 528.06& 196.73& 176.28 &307.17 & 421.26      & 171.95    & 0.27 &0.560 \\
Cal.\cite{becquart2007ab} & 3.17 & 536.32& 202.25& 138.70 &313.61 & 386.83      & 149.42    & 0.29 &0.478 \\
Exp.\cite{soderlind1993theory}&3.16&533&205&163 &314.33 & 417.80      & 163.40    & 0.28 &0.54 \\
\Hline
\end{tabular}
\end{table}
\section{Results and discussion }
\subsection{Optimization of the crystal structure}
\begin{figure}
\includegraphics[width=0.75\textwidth]{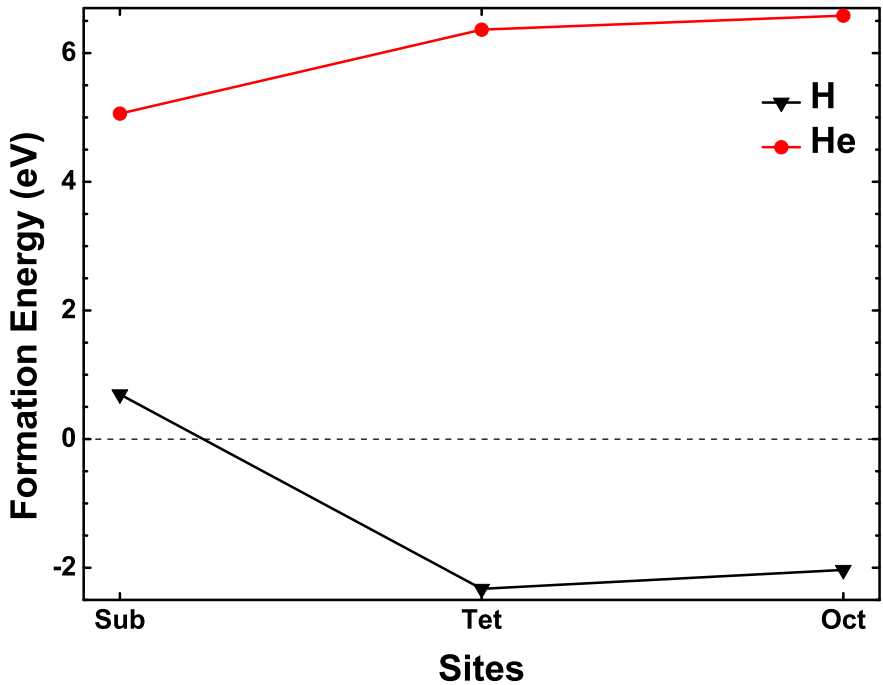}
\caption{The formation energy of the single defects in W.}
\label{fig.4}
\end{figure}
In order to study various defects formed by H or He atoms in W, first of all, we  optimized the crystal structure of the perfect W supercell with and without the various defects formed by H or He atoms. Table 1 shows the calculated results, lattice parameters, elastic constants and elastic modulus of the BCC W. Table 1 shows that our results are in good agreement with the theoretical simulation results\cite{becquart2007ab} and the experimental data\cite{soderlind1993theory}. Fig. 3 shows the change of the volume of BCC W when the H or He atoms was introduced in W. From Fig. 3, we can find that the volume of W decreases when the H or He atoms are located in the substitutional sites while the volume increases when the H or He atoms are located in the tetrahedral and octahedral sites. The reason for the volume decrease is that the atomic radius of H or He are much smaller than that of W, and the volume increase due to the extra atoms that enter the W supercell.
 \begin{figure}
 \includegraphics[width=0.75\textwidth]{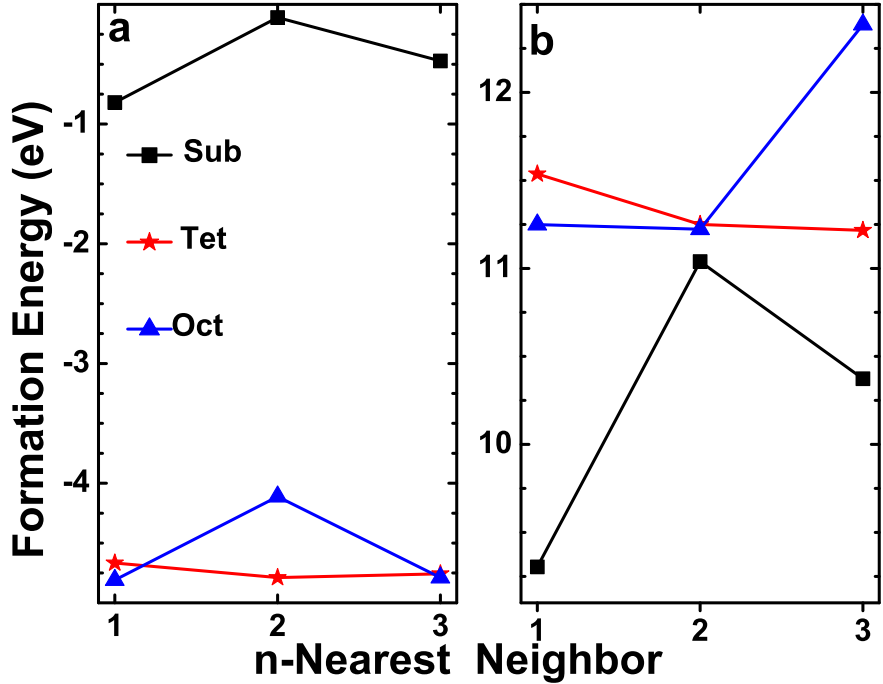}
\caption{The formation energy of two H and He atoms located in the different sites. (a) and (b) show the formation energy two H and He atoms that located in different sites, respectively.}
\label{fig.5}
 \end{figure}
\subsection{Formation energy}
The formation energy of various defect configurations, $E_{\rm{defct}}^{\rm{f}}$, was define as follows:
\begin{equation}
E_{\rm{defect}}^{\rm{f}}=E_{\rm{system}}^{\rm{tot}}-E_{\rm{W}}^{\rm{24}} \left( \frac{N}{24}\right)-n_{\rm{He}}E_{\rm{He}}-n_{\rm{H}}E_{\rm{H}}
\end{equation}
where, the $E_{\rm{system}}^{\rm{tot}}$ is the total energy of the system with defects, $N$, $n_{\rm{He}}$,
 and $n_{\rm{H}}$ are the numbers of W, He and H atoms, respectively, $E_{\rm{W}}^{\rm{24}}$ is the total energy of a perfect $2\times3\times2$ supercell of BCC W with 24 atoms, and $E_{\rm{He}}$ is the energy of an isolated He atom, $E_{\rm{H}}$ is the total energy of an isolated H atom. The formation energy of various single point defects in W are summarized in Table 2, and the results are in good agreement with other calculation results\cite{lee2009energetics,liu2009structure,becquart2007ab}.
 \begin{figure}
\includegraphics[width=0.75\textwidth]{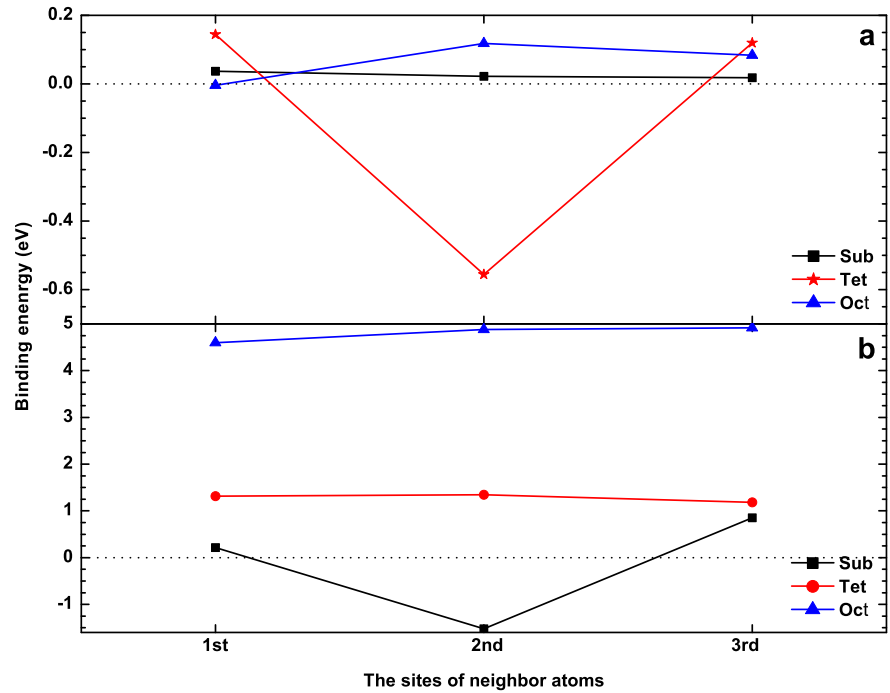}
\caption{The binding energy of two H and He atoms in W when they are located in different positions and the distance of two H and He atoms are increased.}
 \label{fig.6}
 \end{figure}
  According to Table 2, we plotted the formation energies of the single defect in Fig. 4. For He, the substitutional site is the most favorable over the tetrahedral and octahedral sites. For H, the tetrahedral site is the most preferred site over the substitutional and octahedral sites. Interestingly, the formation energies of H$_{\rm{Tet}}$ and H$_{\rm{Oct}}$ are negative, which implies that defects are more stable than other defects, and the formation process of the defects, H$_{\rm{Tet}}$ and H$_{\rm{Oct}}$, is an exothermic process. Fig. 5 shows the formation energy of two H and He atoms in W, and these two atoms are located in different sites. Fig. 5(a) shows the formation energy of two H atoms that are located in different substitution, tetrahedral, and octahedral sites. According to the distance between two H atoms, they are marked as the first, second and third nearest neighbor atoms. Fig. 5(b) shows the formation energy of two He atoms located in different substitutional, tetrahedral, and octahedral sites. For two H atoms, the highest formation energy is the H atoms located in different substitutional sites, and the lowest formation energy is closely related to the sites of the two H atoms. The formation energies of the two H atoms located in different octahedral sites are the lowest when the two H atoms are the first or third nearest neighbor atoms, and the formation energy of the two H atoms that located in different tetrahedral sites is the lowest, and the two H atoms are the second nearest neighbor atoms. For He atoms, the lowest formation energy is the two He atoms located in different substitutional sites. The formation energy of two He atoms located in different tetrahedral sites is the highest when the two He atoms are the first or second nearest neighbor atoms. When they are the third nearest neighbor atoms, the highest formation energy is the two He atoms located in different octahedral sites. The formation energy of the two foreign atoms (H or He) is not equal to twice the value of the single atom, and the formation energy of the two foreign atoms is smaller than two times the value of the formation energy of a single atom. In a short, the formation energy of two H and He atoms are determined by their located sites.
\begin{table}
\caption{Summary of the formation energy of the single defect in W.}
\label{t2}
\begin{tabular}{ccccccc}
\Hline
Configuration &H$_{\rm{Sub}}$ &H$_{\rm{Tet}}$ &H$_{\rm{Oct}}$ &He$_{\rm{Sub}}$ &He$_{\rm{Tet}}$&He$_{\rm{Oct}}$\\
\Hline
 This paper    &0.965 & -2.327  & -2.034  & 5.059  &6.365 &6.583       \\
 Cal. \cite{lee2009energetics}          &0.92  & -2.47   & -2.07   & 5.00   &6.23  &6.48         \\
 Cal. \cite{liu2009structure,becquart2007ab}          &0.78  & -2.44   & -2.06   & 4.70   &6.16  &6.38          \\
\Hline
\end{tabular}
\end{table}
\subsection{Interaction of H-H and He-He}
The binding energy between atoms can reflect the interaction that exists among these atoms. The binding energy of two H or He atom is calculated as follows:
\begin{equation}
E_b(X_2)=2E(X)-E(X_2)-E_W^{24}
\end{equation}
where, $E(X)$ is the total energy of the W supercell with a H or He atom, $E(X_2)$ is the total energy of the supercell with two H or He atoms, and $E_W^{24}$ is the total energy of 24 W atoms. A positive binding energy indicates that attraction exists between the atoms, while a negative binding energy shows that a repulsion interaction exists between the atoms. The binding energies of two H or He atoms are shown in Fig. 6. The most binding energy exists among two H atoms are near the zero while the most binding energy exists among two He atoms are in a positive value. The results show that there is an attraction interaction among the He atoms, and the interaction between two H atoms is very weak.

\begin{figure}
\includegraphics[width=0.75\textwidth]{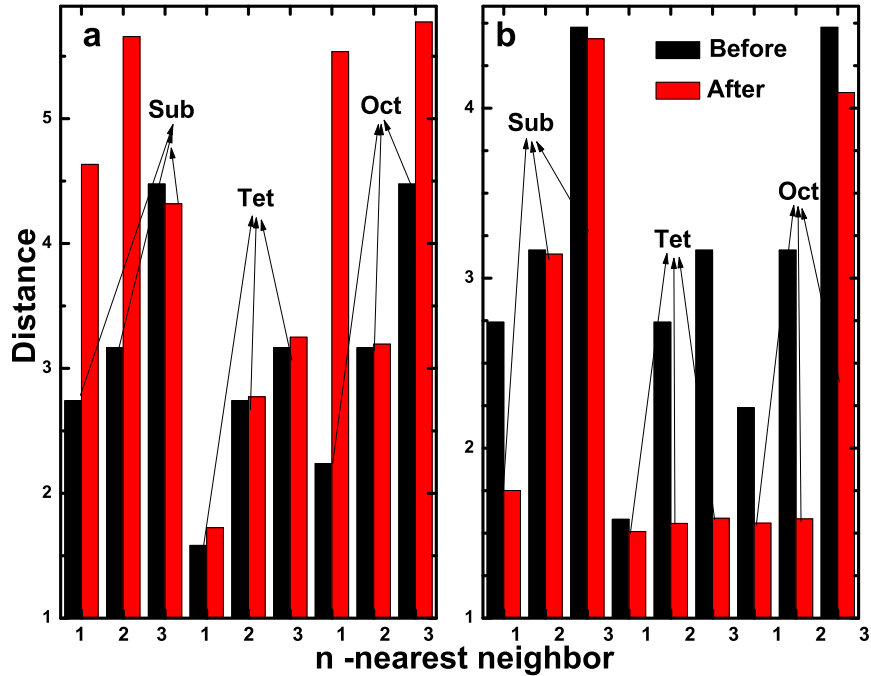}
\caption{The distance (in \AA) between the two H or He atoms before the relaxation and after relaxation. (a) and (b) show the distances between two H and He atoms, respectively.}
\label{fig.7}
\end{figure}

To further study the interaction among two H and He atoms, we recorded the distance between two H and He atoms before and after relaxation. Fig. 7 shows the change of the distance between two H or He atoms after relaxation, and the atoms are located in different sites. The distance between two H atoms becomes longer after relaxation than before relaxation. This result suggests that a repulsive interaction exists between H atoms when the H atoms are close to each other because each H atom in W lattice carries certain amount of negative charge. The distance between two H atoms is larger than 0.75\AA (the distance between two H atoms in a H$_2$ molecule) after relaxation, and no H$_2$ molecule formed. The result of our study is in good agreement with the results of Liu et al.\cite{liu2009structure} and Xu et al. \cite{xu2009first}. The distance between two He atoms becomes shorter after relaxation. This result indicates that there is an attraction interaction between the two He atoms when they close to each other.

\begin{figure}
\includegraphics[width=0.75\textwidth]{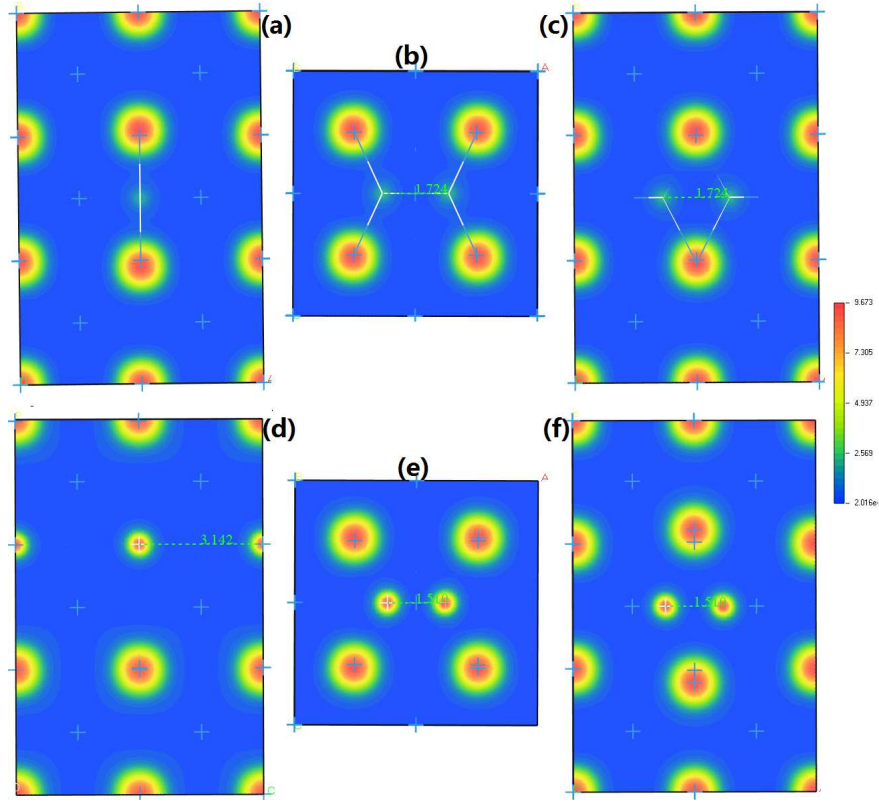}
\caption{(Color online) Charge density distribution. (a), (b), and (c) show the single H atom and two H atoms located in octahedral sites, (d), (e), and (f) show the single He atom and two He atoms located in different octahedral sites. The red color represents high charge density, and the blue color represents low valence charge density.}
\label{fig.8}
\end{figure}

\subsection{Analysis of the Mulliken population and charge density}

\begin{table}
\caption{Atomic populations (Mulliken) of H and He atoms in W}
\label{t3}
\begin{tabular}{ccccc}
\Hline
Sites & Species & Ion & $s$&Charge($e$)\\
\Hline
\multirow{2}{1cm}{H$_{\rm{Sub}}^{21}$}    &H & 1  & 1.10  &-0.10        \\
         &H & 2  & 1.10 &-0.10     \\

\multirow{2}{1cm}{He$_{\rm{Sub}}^{21}$}          &He & 1      & 1.95    &0.05     \\
            &He & 2      & 1.95    &0.05\\
\multirow{2}{1cm}{H$_{\rm{Tet}}^{21}$}          &H & 1  & 1.24  &-0.24    \\
            &H & 2  & 1.10 &-0.24 \\
\multirow{2}{1cm}{He$_{\rm{Tet}}^{21}$}          &He & 1      & 1.89    &0.11     \\
            &He & 2      & 1.89    &0.11 \\
\multirow{2}{1cm}{H$_{\rm{Oct}}^{21}$}          &H & 1  & 1.26  &-0.26     \\
            &H & 2  & 1.26  &-0.26\\
\multirow{2}{1cm}{He$_{\rm{Oct}}^{21}$}          &He & 1      & 1.89    &0.11     \\
            &He & 2      & 1.89    &0.11\\
\Hline
\end{tabular}
\end{table}

The electronic structure can help us understand the root causes of the interaction between two H and He atoms, so the Mulliken population and the charge density were analyzed in this paper. Table 3 shows the atomic population of the H and He atoms in W, and two H and He atoms are the first nearest neighbor atoms. Mulliken population analysis shows that certain electrons will transfer from W to H, and H atoms can stand in a stable state. The H atoms will accept 0.10 electrons from the surrounding W atoms when H atoms are located in substitutional sites. The H atoms accept 0.24 electrons when the H atoms are located in the tetrahedral sites, and they accept 0.26 electrons when they are located in the octahedral sites. The amount of charge transfer is insensitive to the interstitial H and He atoms' positions. The H atoms have a negative charge. This implies that there is Coulomb repulsion between two H atoms. On the contrary, the W will get some electrons from the He atoms. The W atoms will accept 0.05 electrons from a He atom when the He atoms are located in the substitution sites, W atoms will get 0.11 electrons from a He atom when He atoms are located in the tetrahedral and octahedral sites. Thus, the Van der Waals force between two He atoms becomes stronger when they are close to each other. The more electrons the W atoms get, more stable the structure is. Interestingly, we found that the H-H repulsive interaction fit well to the screened Coulomb potential \cite{kittel1986introduction}. To explain the differences between the interactions between two H atoms and He atoms, charge density distribution was used. We selected a single atom and two atoms located in the octahedral site. Fig. 8 shows the charge density of one and two atoms. Fig. 8(a) and (d) show the charge density of an H and He atom in W, respectively. Fig. 8(b) and (c) show two H atoms in W, and Fig. 8(e) and (f) also show two H atoms in W from several different perspectives. The bonds between H and W contain covalent components. The bond is stronger than that between He and W, and the charge density difference between H and W is bigger than that between He and W. The H atoms more likely to be attracted by W atoms than by each other while the He atoms are likely to attract each other. There is only one electron in the 1$s$ orbital of an H atom, and the H atom obtains the electrons from the W atoms to maintain a relatively stable state. An He atom is a closed shell atom. It prefers to occupy the low electron density zone (interstitial sites).

\section{Conclusion}
Using first-principles calculation, we studied the BCC W with single and two point defects formed by H or He
atoms, respectively. The results of our calculations are as follows:

(1)For a single H atom, the most preferred site is the tetrahedral interstitial site. For He, the substitutional site is more favorable than the tetrahedral and octahedral sites.

(2)For two H and He atoms, the formation energy is closely related to the sites of the two atoms. The highest formation energy of H atoms was found in the two H atoms that were located in different substitutional sites, while the lowest formation energy of He atoms was found in the two He atoms located in different substitutional sites.

(3)After relaxation, two H atoms are going away from each other because the H atom accepts some charge from the surrounding W atoms and they form a covalent bond. The He atoms close to each other because the W atoms get some electrons from the He atoms, and the more electrons the W atoms get, the more stable the crystal stucture is.

 Our results are good in agreement with other simulation results and experimental data. This paper
 might shed some light on the development of tungsten materials used in ITER and other fusion reactors.
 
 \section*{Conflicts of Interest}
 
 The authors declare that there is no conflict of interest regarding the publication of this paper.

\acknowledgment
This work was supported by the National Natural Science Foundation of China under Grant Nos. 11275071 and 11305061,
and the Fundamental Research Funds for the Central Universities under Grant No. 2014MS53.

 \end{document}